\documentclass[
  aps,
  prl,
  twocolumn,
  groupedaddress,
  longbibliography
]{revtex4-2}
\usepackage{graphicx}
\usepackage[dvipsnames]{xcolor}
\usepackage{amsmath}
\usepackage{float}
\definecolor{darkred}{rgb}{0.80,0,0}
\definecolor{blood}{rgb}{0.50,0,0}
\definecolor{brightred}{rgb}{1,0,0}
\definecolor{orange}{rgb}{1,0.3,0}
\definecolor{bluegreen}{rgb}{0,0.5,0.5}
\definecolor{lightblue}{rgb}{0,0.5,0.8}
\definecolor{darkgreen}{rgb}{0,0.5,0}
\definecolor{green}{rgb}{0,0.70,0}
\definecolor{darkblue}{rgb}{0,0,0.80}
\definecolor{magenta}{rgb}{1,0,1}
\definecolor{softmagenta}{rgb}{0.85,0.1,0.6}
\definecolor{mauve}{rgb}{0.6,0.1,1}
\definecolor{white}{rgb}{1,1,1}
\definecolor{black}{rgb}{0,0,0}

\usepackage[
    colorlinks=true,
    linkcolor=blue,
    citecolor=blue,
    urlcolor=blue]{hyperref}

\emergencystretch=\maxdimen
\begin{document}
\title{Nonmonotonic Magnetic Correlations in Decorated Lieb-Lattice Anderson Models}

\author{R.~Hmuifoo}
\affiliation{Department of Physics and Astronomy
University of Notre Dame,
Notre Dame, IN 46556 USA} 

\author{E. W.~Huang}
\affiliation{Department of Physics and Astronomy,
University of Notre Dame,
Notre Dame, IN 46556 USA} 
\affiliation{Stavropoulos Center for Complex Quantum Matter, University of Notre Dame, Notre Dame, IN, USA}

\author{A.~Ancheta}
\affiliation{Department of Physics and Astronomy, One Shields Ave.,
University of California, Davis, California 95616, USA}

\author{R. T.~Scalettar}
\affiliation{Department of Physics and Astronomy, One Shields Ave.,
University of California, Davis, California 95616, USA}

\begin{abstract}
The interplay between electronic correlations and lattice geometry can strongly
modify the competition between magnetic correlations and local singlet formation
in multiorbital Hamiltonians. Here we use determinant quantum Monte Carlo
simulations to investigate two periodic Anderson-like models constructed by
selectively hybridizing localized orbitals to the inequivalent sites of a
Lieb lattice. The resulting geometries possess partially filled flat bands
and different sublattice imbalances. We examine the evolution of local
spin correlations and orbital-resolved magnetic structure factors as the
hybridization is increased. Although correlations between hybridized
orbitals evolve relatively smoothly toward local singlet formation, the
longer-range magnetic correlations exhibit a striking nonmonotonic dependence
on hybridization. In particular, we observe a pronounced suppression at
intermediate hybridization followed by a substantial recovery on undecorated
lattice sites. We discuss these results in relation to earlier studies
of depleted periodic Anderson models and the constraints imposed by Lieb's
theorem. Our findings illustrate how selective hybridization and flat-band
structure can produce unusual redistributions of magnetic correlations in
strongly interacting multiorbital systems.
\end{abstract}

\maketitle

\section{I.~Introduction}

The single band Hubbard model was originally proposed to understand magnetic and insulating behavior caused by electron-electron interactions~\cite{Hubbard1963,Gutzwiller1963,Kanamori1963},
but has proven to have a surprisingly eclectic set of applications
both in the condensed matter~\cite{white89,rasetti91,montorsi92,scalapino1994,gebhard97,fazekas99,Lee2006,Vojta2009,Keimer2015,Greene2020,Arovas2022}
and ultracold atom communities~\cite{Hofstetter2002,Bloch2008,Mazurenko2017}.
The inclusion of multiple bands has also proven essential for modeling related many-body phenomena.  
Important examples include the Kondo and Anderson models of magnetic impurities
in solids and their extensions to dense arrays which serve as descriptions
of heavy fermion materials and especially the competition of antiferromagnetic order such as occurs in 
CeAl$_2$ and the non-magnetic (singlet) ground states of materials like CeAl$_3$~\cite{Stewart1984,Tsunetsugu1997,Riseborough2000,Coleman2007,Gegenwart2008,Coleman2015}.

Although the focus of strongly correlated physics in such models is often on the on-site repulsion $U$,
the Kondo and PAM cases noted above emphasize that the lattice geometry (in a real space picture), or equivalently
the band structure (in momentum space), can affect the nature of the low temperature phases.  Frustrated lattices offer another illustration of the sensitivity
of magnetism to geometry: on triangular and kagome lattices, competing
correlations can suppress conventional antiferromagnetic order
\cite{Yoshioka2009,Szasz2020,Wietek2021,Ohashi2006,Sun2021,Kaufmann2021,Yan2011}.
The present work instead concerns \emph{bipartite}, unfrustrated lattices
with an imbalance of sites between the two sublattices.

Even in the absence of frustration, bipartite lattices can have
fundamental differences.
In the case of the single band Hubbard model, a square lattice at half-filling
has perfect nesting as well as a van-Hove singularity in the density of states, and
hence any non-zero interaction strength $U > 0$ results in 
an antiferromagnetically  ordered insulator~\cite{Hirsch1985,white89,Schulz1990PRL,CiniStefanucci2001}.
This is in contrast to, for example, the honeycomb lattice where a finite critical $U_{c} \sim 3.87$
(in units of the hopping $t$) must be exceeded to exit the metallic phase~\cite{Paiva2005,Meng2010,Sorella2012}.
In such two dimensional lattices true long range order 
can only occur at $T=0$, owing to the continuous nature of the spin
order parameter~\cite{MerminWagner1966,Hohenberg1967,Mermin1968,KosterlitzThouless1973}.

Another interesting configuration of a bipartite lattice is one with
a partially filled flat band at the Fermi energy. This occurs for the
Lieb lattice, where a sublattice imbalance produces a zero-energy flat
band. For a connected half-filled bipartite Hubbard model with
repulsive onsite interactions, Lieb's theorem fixes the ground-state
total spin at $S_{\rm GS}=|N_{\cal A}-N_{\cal B}|/2$; when the
sublattices are unequal, this implies a ferrimagnetic ground-state
spin rather than a total-spin singlet~\cite{Lieb1989,Costa2016}.
Two distinct cases occur here depending on whether or not the flat band is separated by gaps from the dispersive bands.

The effects of flat bands on the competition
between magnetic order and singlet behavior have been examined in several computational studies.
For a periodic Anderson model on a square lattice, where half of the localized $f$ orbitals 
are removed from one of the sublattices, the behavior of the local moments and compressibility
were shown to exhibit 
distinct behaviors for the conduction orbitals with and without local orbital partners~\cite{Costa2018}.
For a PAM on a Lieb-lattice conduction backbone, with a different
pattern of local-orbital attachment and interactions, the competition
between magnetic correlations and local singlet formation was
investigated~\cite{Hartman2016}.
Finally, work on such 2D lattices was extended to a fully
3D geometry of multi-layer Lieb lattices in order to locate
the finite transition temperatures $T_c$ to magnetic order~\cite{Noda2015}.  In this
case a key focus was on understanding the effects of the flat band on the
functional dependence of $T_c$ on the on-site interaction $U$, generalizing what
was known about the effect of a (square lattice) van Hove singularity in  the density
of states.

Recent experiments have also begun to explore the magnetic properties of Lieb lattices. Ferrimagnetic correlations have been observed in ultracold fermions on a Lieb lattice~\cite{Lebrat2026Ferrimagnetism}. Cluster-decorated Lieb lattices have also been constructed and their flat band structure examined using scanning tunneling spectroscopy~\cite{Li2026ClusterDecorated}. Similar Lieb-like electronic structures have been investigated in covalent-organic frameworks~\cite{Cui2020LiebCOF}. Although these systems do not share the exact Hamiltonians considered here, they demonstrate the growing experimental attention to Lieb-lattice geometries and their electronic properties.

Another related area is the competition between local hybridization and magnetic correlations in heavy fermion materials. For example, orbital-selective Kondo behavior has been seen in USb$_2$~\cite{Giannakis2019OrbitalSelective}, where it coexists with antiferromagnetism.  This provides an experimental example of the interaction between local hybridization and magnetic order.

The work reported in this paper builds upon these studies.
Specifically, we combine the Lieb-lattice conduction backbone
of Ref.~\cite{Hartman2016} with selective attachment of local
orbitals, in the spirit of Ref.~\cite{Costa2018}. We compare attaching
the $f$ orbitals to the fourfold-coordinated $d$ sites with attaching
them to the twofold-coordinated $p$ sites. Unlike the conventional PAM,
we also retain a Hubbard repulsion on the conduction backbone. These
two choices produce different flat-band spectra and sublattice
imbalances. The distinctive observation is a pronounced minimum in
orbital-resolved magnetic correlations at intermediate hybridization,
followed by their recovery on undecorated backbone orbitals.
Because the exact ground-state total spin remains nonzero,
we interpret this minimum as a suppression of magnetic correlations 
resulting from a reduced magnetic energy scale.

The remainder of this paper is organized as follows:
In Section II, we discuss our model, its non-interacting limit, the
quantum Monte Carlo methodology, and the observables measured.
In Section III, we present the magnetic correlations and their
nonmonotonic dependence on hybridization. Section IV summarizes the results.

\section{II.~Model and Methods}

We consider a Lieb lattice consisting of a square array
of $d$ sites with two symmetric bridging $p$ orbitals
along the horizontal and vertical bonds. The kinetic energy is,
\begin{align}
    \hat {\cal K}_{\rm Lieb} = -t \sum\limits_{i,\sigma} \, \Big(
&\,\hat d^{\dagger}_{i,\sigma} \big( \hat p_{i,x,\sigma}^{\vphantom{dagger}}
+ \hat p_{i,y,\sigma}^{\vphantom{dagger}} \big) 
\nonumber
\\ 
+&\,\hat p^{\dagger}_{i,x,\sigma}\hat d_{i+\hat x,\sigma}^{\vphantom{dagger}}
+\hat p^{\dagger}_{i,y,\sigma}\hat d_{i+\hat y,\sigma}^{\vphantom{dagger}} \, \Big)+ {\rm h.c.}
\label{eq:KLieb}
\end{align}
Here $t$ is the hybridization between $d$ orbitals with 	cb{creation (annihilation) operators} 
$\hat d_{i,\sigma}^{\dagger} (\hat d_{i,\sigma}^{\phantom{\dagger}})$
and $p$ orbitals with 	cb{creation (annihilation) operators} 
$\hat p_{i,\sigma}^{\dagger} (\hat p_{i,\sigma}^{\phantom{\dagger}})$.
The sum is over two spin species $\sigma = \uparrow,\downarrow$.
We define $L$ to be the number of $d$ sites in each of the $x$
and $y$ directions, so that there are $N=L \times L$ unit cells in the Lieb lattice.

We also hybridize local $f$ orbitals to selected sites of this Lieb-lattice `backbone'.
We consider two models.  
In the first, the $f$ orbitals attach solely to the $d$ sites of the Lieb backbone,
\begin{align}
    \hat {\cal K}_{{\rm loc},d} = -t^{\prime} \sum\limits_{i,\sigma} \, ( \,
 \hat d^{\dagger}_{i,\sigma} \hat f_{i,\sigma}^{\vphantom{dagger}} 
 + \hat f^{\dagger}_{i,\sigma} \hat d_{i,\sigma}^{\vphantom{dagger}}  \, )
\label{eq:Kdecord}
\end{align}
In the second, the $f$ orbitals are present only on the Lieb backbone $p$ sites,
\begin{align}
    \hat {\cal K}_{{\rm loc},p} = -t^{\prime} \sum\limits_{i,\sigma} \, \big(
 \, \hat p^{\dagger}_{i,x,\sigma} \hat f_{i,x,\sigma}^{\vphantom{dagger}} 
 + \hat p^{\dagger}_{i,y,\sigma} \hat f_{i,y,\sigma}^{\vphantom{dagger}} + {\rm h.c.} \,\big) \,.
\label{eq:Kdecorp}
\end{align}
In both cases, $t^{\,\prime}$ denotes the hybridization to the localized orbitals.
We set $t=1$ as our scale of energy and study the magnetic
correlations as they vary with $t^{\prime}$.
The horizontal and vertical $p$ orbitals are related by lattice symmetry,
as are the corresponding local $f$ orbitals in
$\hat{\cal K}_{{\rm loc},p}$.
We therefore suppress the $x$ and $y$ labels when discussing
symmetry-equivalent observables.

\begin{figure}[t!]
\includegraphics[height=5.0cm,width=5.0cm,angle=0]{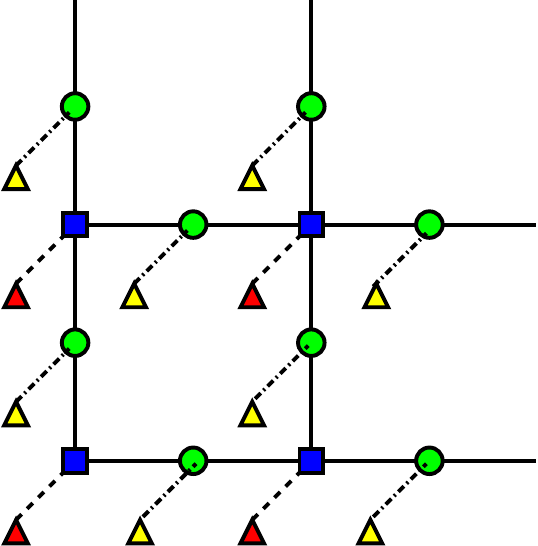}
\caption{
The real space lattice geometries considered in this work.
The Lieb lattice backbone consists of $d$ orbitals (blue squares) and two
symmetry-equivalent $p$ orbitals (green circles).
For $\hat{\cal K}_{{\rm loc},d}$, a local $f$ orbital (red triangle) is
attached to each $d$ orbital, giving four orbitals per unit cell.
For $\hat{\cal K}_{{\rm loc},p}$, local $f$ orbitals (yellow triangles) are
attached to the two $p$ orbitals, giving five orbitals per unit cell.
The two $p$ orbitals are related by lattice symmetry, as are the two $f$
orbitals in $\hat{\cal K}_{{\rm loc},p}$.
Accordingly, throughout this work we use $p$ and $f$ without $x$ or $y$
subscripts when referring to symmetry equivalent observables --
$c_{pp}(r)=c_{p_xp_x}(r)=c_{p_yp_y}(r)$ and
$c_{ff}(r)=c_{f_xf_x}(r)=c_{f_yf_y}(r)$.
(See Eq.~\ref{eq:spincorr}.)
Solid bonds denote hopping $t$, and dashed bonds denote hybridization
$t^{\prime}$.
We show $N=L\times L=2\times2$ unit cells.
\label{fig:geom}
}
\end{figure}

\begin{figure}[t!]
\centering
\includegraphics[width=\linewidth]{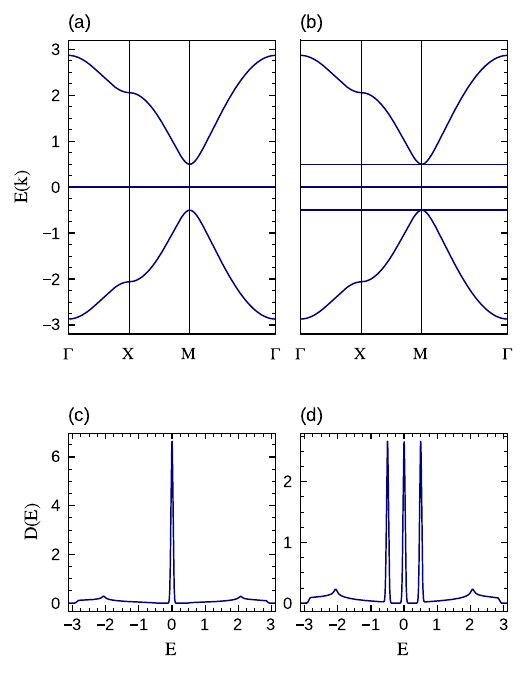}
\caption{
Panels (a) and (b) show the band structures 
of the four band Lieb lattice  with  local orbitals on the $d$ sites and 
of the five band Lieb lattice  with  local orbitals on the $p$ sites, respectively.
The hopping to the localized orbitals is $t^{\prime}=0.5$.
Panels (c) and (d) show the corresponding densities of states.
The two dispersive bands are identical in the two geometries.  The difference is that for
$\hat {\cal H}_{{\rm loc},d}$
there are two flat bands at the half-filled Fermi energy, whereas
$\hat {\cal H}_{{\rm loc},p}$
has  three flat bands at $E(k_{x},k_{y}) = -t^{\prime}, 0, +t^{\prime}$.
\label{fig:band}
}
\end{figure}

Figure \ref{fig:geom} gives a picture of these two geometries.
$\hat {\cal K}_{\rm Lieb}$ encompasses the blue squares and green circles,
while
$\hat {\cal K}_{\rm loc,d}$
and
$\hat {\cal K}_{\rm loc,p}$
are the `localized' red or yellow triangles.
The associated band structures of 
$\hat {\cal K}_{\rm Lieb} + \hat {\cal K}_{\rm loc,d}$
and
$\hat {\cal K}_{\rm Lieb} + \hat {\cal K}_{\rm loc,p}$,
along with their densities of states, are shown in Fig.~\ref{fig:band}.
Writing $\varepsilon_{\boldsymbol{k}}^2 =
4t^2[\cos^2(k_x/2)+\cos^2(k_y/2)]$, the noninteracting
four-orbital model has two flat bands at $E=0$ and two dispersive
bands $E=\pm\sqrt{\varepsilon_{\boldsymbol{k}}^2+t'^2}$.
The five-orbital model has flat bands at $E=0,\pm t'$ and the
same two dispersive bands. Thus the two decorations differ in their
flat-band spectra while sharing identical dispersive branches. For
$t'>0$ the zero-energy flat bands are separated from the dispersive
branches.

Finally, we place a Hubbard $U$ on each orbital,
\begin{align}
  \hat {\cal V} =  \sum\limits_{i} \, U_{d}
 &\, (\hat n_{d,i,\uparrow}-\tfrac{1}{2}) (\hat n_{d,i,\downarrow}-\tfrac{1}{2})
\nonumber
\\ 
 +\, U_{p} & \, (\hat n_{p_{x},i,\uparrow}-\tfrac{1}{2}) (\hat n_{p_{x},i,\downarrow}-\tfrac{1}{2})
\nonumber
\\ 
 +\, U_{p} & \, (\hat n_{p_{y},i,\uparrow}-\tfrac{1}{2}) (\hat n_{p_{y},i,\downarrow}-\tfrac{1}{2}) 
\nonumber
\\ 
 +\, U_{f} &\, (\hat n_{f,i,\uparrow}-\tfrac{1}{2}) (\hat n_{f,i,\downarrow}-\tfrac{1}{2}) 
\label{eq:VLieb}
\end{align}
In Eq.~\ref{eq:VLieb}, the last line is summed over the single
$f$ orbital in each unit cell for $\hat{\cal H}_{{\rm loc},d}$,
and over both $f_x$ and $f_y$ orbitals for
$\hat{\cal H}_{{\rm loc},p}$.
We have written the interaction term 
in `particle-hole' symmetric form.  With this convention, since there are no site energies or overall chemical potential,
these bipartite lattices are both half-filled for all temperatures $T$ and 
Hamiltonian parameters $t, t^{\prime}, U_{\rm d}, U_{\rm p}, U_{\rm f}$.
We will mainly consider the case
$U_{\rm d}= U_{\rm p}= U_{\rm f} \equiv U_{0}$.
Unlike a conventional PAM with a repulsion only on the localized
$f$ orbitals, this choice permits local moments to form directly on
the undecorated conduction sites. Half-filling optimizes the tendency for AF correlations. 
Additionally, there is no sign problem~\cite{loh90,troyer05}, allowing
DQMC simulations at low temperature.

We will refer to our two Hamiltonians as,
$\hat {\cal H}_{{\rm loc},d} = \hat {\cal K} + \hat {\cal K}_{{\rm loc},d} + \hat {\cal V}$
and
$\hat {\cal H}_{{\rm loc},p} = \hat {\cal K} + \hat {\cal K}_{{\rm loc},p} + \hat {\cal V}$.
Our simulations are performed on lattices with a square $L \times L$ array of unit cells, with
four (five) orbitals per unit cell for
$\hat {\cal H}_{{\rm loc},d}  (\hat {\cal H}_{{\rm loc},p} )$.

The properties of these models can be obtained via determinant quantum Monte Carlo (DQMC) simulations.
Here we highlight key aspects of the method.
For details see~\cite{blankenbecler81,white89,AssaadEvertz2008,GubernatisDeterminant2016}.
In DQMC a path integral is constructed for the partition function
${\cal Z} = {\rm Tr} e^{-\beta \hat {\cal H}}$
by discretizing the imaginary time $\beta$ into $L_\tau=\beta/\Delta\tau$ intervals of length $\Delta \tau$.
This allows the use of the Trotter approximation
$e^{-\Delta \tau \hat {\cal H}}
 \approx  e^{-\Delta \tau \hat {\cal K}}
   e^{-\Delta \tau \hat {\cal V}}$ and subsequently the
  Hubbard-Stratonovich transformation to convert
  $  e^{-\Delta \tau \hat {\cal V}}$ into a quadratic form of fermion operators coupled to
  an auxiliary field.  
  At this point, the trace over the fermionic degrees of freedom can
  be performed analytically, leaving ${\cal Z}$ as a sum over the auxiliary field with
  a weight proportional to the product of two determinants, one for each spin species.
  Because of the particle-hole symmetry, these two determinants have the same sign
  regardless of the field configuration or the parameters in the Hamiltonian.
In this work we choose $\Delta \tau = 0.05$, a value for which the systematic
Trotter errors are of the same order, or less than, statistical errors from the Monte Carlo sampling.

The dimension $d_M$ of the matrices $M_\sigma$ whose determinants give the Boltzmann weight is
given by the number of spatial sites in the lattice, which is the number of orbitals 
times the number of unit cells $N$.  For ${\cal \hat H}_{\rm loc,d}$
and ${\cal \hat H}_{\rm loc,p}$ this dimension is $d_M = 4N$ and $d_M = 5N$ respectively.
The computational cost of DQMC scales as $d_{M}^{\,3}$. 
The elements of the inverses $G_\sigma = M_\sigma^{-1}$, when averaged by the Monte Carlo
over Hubbard-Stratonovich field configurations, give the single-particle Green functions
$ \langle \hat c^{\phantom{\dagger}}_{i\sigma} \hat c^{\dagger}_{j \sigma} \rangle$ 
for $\hat c = \hat d, \hat p, \hat f$.
Measurements of observables involving four fermion operators such as the double occupancy or
spin-spin correlations are obtained from products of elements of $G_{\sigma}$ as determined
by Wick's theorem.

The magnetic properties of the PAM
are characterized by intra- and inter-orbital real space spin-spin correlations,
\begin{align}
c^{\phantom{\dagger}}_{\rm \alpha\beta}(r) &= 
\langle \, (n_{\alpha,i+r \, \uparrow} -n_{\alpha,i+r \, \downarrow})
(n_{\beta,i \, \uparrow} -n_{\beta,i \, \downarrow}) \, \rangle
\label{eq:spincorr}
\end{align}
Here the orbital indices $\alpha,\beta$ will run over both the $d,p$ sites in
the Lieb backbone and the $f$ impurities.  $r$ is the spatial separation of the
$z$ components of the spin.  The expression in Eq.~\ref{eq:spincorr} involves the $z$ component of spin. By symmetry, the $x$ and $y$ correlators are identical
to within statistical error bars.  

The $r=0$ correlators play a prominent role in
the physics.  
When $\alpha=\beta$, the onsite value of $c_{\alpha \alpha}(r=0)$ is
the local moment $\langle \, m^2_{\alpha i}\, \rangle = \langle \, (n_{\alpha, i\uparrow} - n_{\alpha , i\downarrow})^2\, \rangle$.  
When $\alpha=d,p$ and $\beta=f$, the onsite value of $c_{\alpha \beta}(r=0)$ measures the
degree of singlet formation between the conduction ($\alpha=d,p$) and localized ($\beta=f$) orbitals.
In correspondence with the usual arguments for the single band Hubbard model, the first effect of $U$
as the temperature $T$ is lowered
is to minimize the potential energy by suppressing double occupancy.  
$ D_{\alpha i}  = n_{\alpha, i \, \uparrow}  n_{\alpha , i \, \downarrow}$.  
This can equivalently
be viewed as promoting moment formation since
$\langle m^2_{\alpha  \, i}\rangle = \langle \, n_{\alpha, i \, \uparrow} + n_{\alpha , i\, \downarrow}  
- 2 D_{\alpha, i}   \, \rangle$.  

Second, as the temperature is lowered further below the exchange energy scale
$J \sim t^2/U$, antiparallel alignment of fermion spins on adjacent spatial sites occurs.
This second effect is driven 	cb{by} the minimization of the kinetic energy by the virtual hopping which
occurs for fermions of opposite spins, but which is forbidden by the Pauli principle for like spins.
At sufficiently low $T$ this local anti-alignment can extend to larger spatial separation,
resulting in long range magnetic order.  

Long range order can be inferred either by looking at the opposite extreme to $r=0$, namely maximal separation
$r=(L/2,L/2)$ on the spatial lattice with periodic boundary conditions, or via the structure factor,
\begin{align}
S_{\alpha \alpha}(q) = \sum_r e^{i q \cdot r} c^{\phantom{\dagger}}_{\rm \alpha\alpha}(r)
\label{eq:structurefactor}
\end{align}
Here $q=(q_x,q_y)$.  
When the order is short ranged $c_{\alpha\alpha}(r)$ will fall off exponentially
and the sum over $r$ in Eq.~\ref{eq:structurefactor} will get contributions only over 
a region $r<\xi$.  Once the linear lattice size $L\gg\xi$, the structure
factor becomes approximately independent of lattice size.  On the other hand, 
when the order is long range 
$S_{\alpha \alpha}(q) \sim N$, for an appropriately chosen wave vector $q$.
Although we show results for $c_{\alpha \beta}(r)$ for $\alpha \neq \beta$ below,
the equal-orbital structure factors provide an orbital-resolved
measure of the spatial extent of magnetic correlations. A
thermodynamic ordered moment requires appropriate finite-size scaling;
a rise with $L$ alone is not conclusive.

For multi-orbital, bipartite geometries such as ours in which the numbers of sites on the
two sublattices are unequal, Lieb emphasized the idea of {\it ferrimagnetic} order.  
Sites belonging to a particular orbital $\alpha$ are
always on the same sublattice and hence have parallel spin (ferromagnetic).
For this reason we focus on $S_{\alpha\alpha}\big(q=(0,0)\big)$.
For $\hat {\cal H}_{{\rm loc},d}$, sites on orbitals $p,f$ share the same sublattice,
and sites on orbital $d$ are on the other sublattice.  
It is worth emphasizing that there
are $p$ orbitals on both the $x$ and $y$ bonds, so that in this case three ($p_x,p_y,f$) of the four orbitals
are on one sublattice and one ($d$) on the other.
Meanwhile, for  $\hat {\cal H}_{{\rm loc},p}$
sites on orbital $p$ share the same sublattice,
and sites on orbitals $d,f$ are on the other sublattice.
Hence three of the five orbitals ($d,f_x,f_y$)
are on one sublattice and two ($p_x,p_y$) are on the other.
We have written $f_x$ and $f_y$ to distinguish local orbitals
attached to $p_x$ and $p_y$.
See Fig.~\ref{fig:geom}.

At half filling and for uniform $U_0>0$, the connected bipartite
Hamiltonians obey Lieb's theorem~\cite{Lieb1989}. With $N=L^2$
unit cells, the four-orbital geometry has $(N_{\cal A},N_{\cal B})
=(N,3N)$ and hence $S_{\rm GS}=N$; the five-orbital geometry has
$(N_{\cal A},N_{\cal B})=(2N,3N)$ and hence $S_{\rm GS}=N/2$.
These nonzero ground-state spins hold for every $t'>0$, even when
individual orbital-resolved magnetic correlations are strongly
suppressed at accessible temperatures. The theorem constrains the
total spin but does not, by itself, determine the finite-temperature
correlation length or the individual orbital structure factors.

\begin{figure}[t!]
\includegraphics[height=4.286cm,width=7.5cm,angle=0]{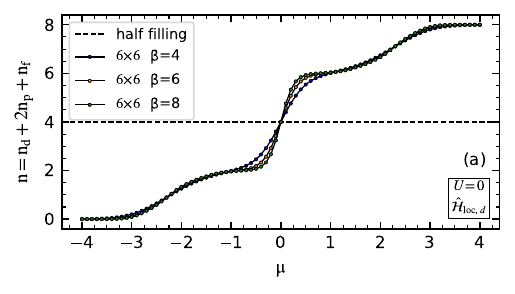}
\includegraphics[height=4.286cm,width=7.5cm,angle=0]{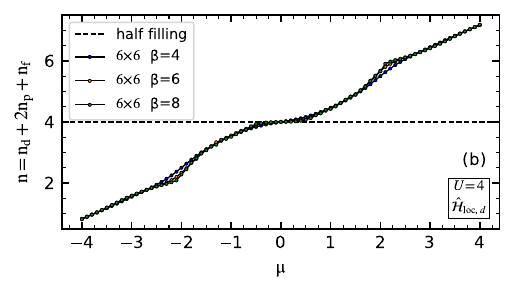}
\includegraphics[height=4.286cm,width=7.5cm,angle=0]{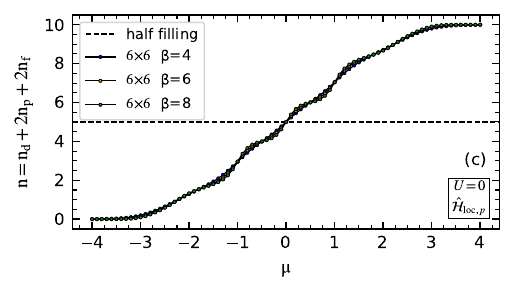} 
\includegraphics[height=4.286cm,width=7.5cm,angle=0]{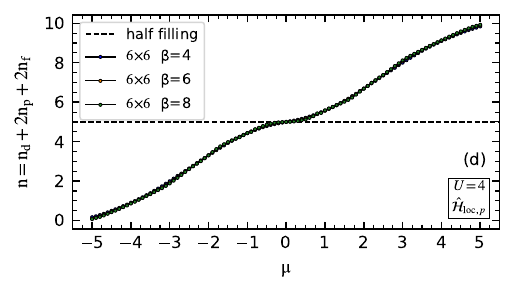} 
\caption{
Density $n(\mu)$ for (a) $U=0$ and (b) $U=4$
at inverse temperatures $\beta=4,6,8$ for
$\hat{\cal H}_{{\rm loc},d}$.
Panels (c) and (d) show the corresponding results for
$\hat{\cal H}_{{\rm loc},p}$.
Lattice size is $6 \times 6$ unit cells, and  
$t^{\prime}=1$.
Half filling occurs at $n=4$ for $\hat{\cal H}_{{\rm loc},d}$ and at $n=5$ for 
$\hat{\cal H}_{{\rm loc},p}$.
At half filling, the $U=0$ flat bands lead to a large
finite-temperature compressibility $\kappa=\partial n/\partial\mu$,
which diverges only in the zero-temperature limit. For $U=4$,
the density plateau signals strongly suppressed compressibility
and interaction-induced insulating behavior.
\label{fig:mottplateau}
}
\end{figure}

\section{III.~Results}

In the absence of interactions, $U_0=0$, the density rises very
steeply at half-filling $\mu=0$ as the flat bands become occupied.
The associated large compressibility, which diverges as
$T\rightarrow0$ in the noninteracting limit, is seen in
Fig.~\ref{fig:mottplateau}(a) for an $L=6$ system with $t'=1$
for $\hat{\cal H}_{{\rm loc},d}$.
When $U_0$ is made nonzero and $\beta$ is sufficiently large,
the rapid density change is replaced by a plateau, indicating
strong suppression of the compressibility and an interaction-induced
charge gap at low temperature, Fig.~\ref{fig:mottplateau}(b).
Results are also shown in panels (c) and (d) for
$\hat{\cal H}_{{\rm loc},p}$. The qualitative behavior is similar,
but differences arise from the flat-band spectra: the two zero-energy
flat bands in $\hat{\cal H}_{{\rm loc},d}$ enhance the half-filled
charge response at $U=0$, whereas the extra flat bands at $\pm t'$
in $\hat{\cal H}_{{\rm loc},p}$ yield large responses near $n=3$
and $n=7$. See Fig.~\ref{fig:band}.

\begin{figure}[t!]
\includegraphics[height=6.0cm,width=7.5cm,angle=0]{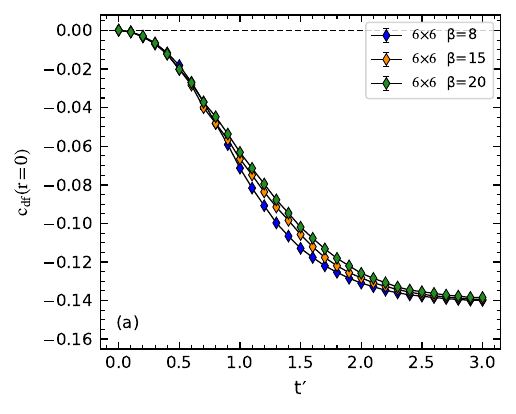}
\includegraphics[height=6.0cm,width=7.5cm,angle=0]{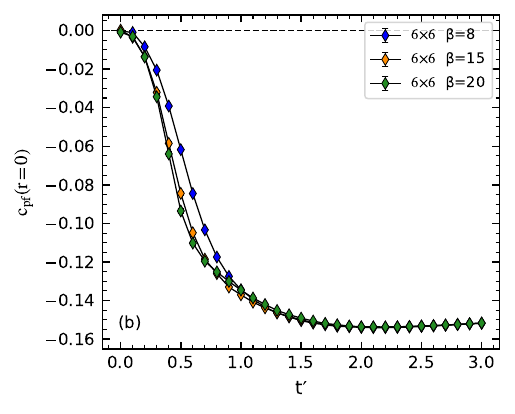}
\caption{
The local singlet correlator as a function of $t^{\prime}$.
Panel (a) shows $c_{df}(r=0)$ for
$\hat{\cal H}_{{\rm loc},d}$, and panel (b) shows
$c_{pf}(r=0)$ for
$\hat{\cal H}_{{\rm loc},p}$.
As $t^{\prime}$ grows, the corresponding backbone and localized
$f$ moments become increasingly anti-correlated.
The lattice size is $6\times6$ unit cells with inverse temperatures
$\beta=8,15,20$.
\label{fig:singlet}
}
\end{figure}

In the half-filled insulating phase different magnetic patterns can occur.  
The top panel of 
Fig.~\ref{fig:singlet} gives the singlet correlator $c_{df}(r=0)$ between $d$ and $f$
for the four orbital geometry of $\hat{\cal H}_{{\rm loc},d}$,
as a function  of the $d$-$f$ hybridization 
$t^{\prime}$.
The correlator $c_{df}(r=0)$ vanishes at
$t^{\prime}=0$ but rapidly becomes more negative as $t'$ grows.
This is consistent with the development of local antiferromagnetic
$d$--$f$ correlations, although the $z$-component correlator alone does
not measure a pure-singlet probability. As shown below, the smooth
growth of this local anticorrelation contrasts with a strongly
nonmonotonic response of longer-distance correlations.
The bottom panel of 
Fig.~\ref{fig:singlet} gives the analogous singlet correlator $c_{pf}(r=0)$ between the $p$ and $f$
sites for the five orbital geometry of $\hat{\cal H}_{{\rm loc},p}$.
Local correlators ($r$ small) are expected to be less sensitive to
linear lattice size $L$ and inverse temperature $\beta$ than their longer range counterparts.
Here $\beta=15$ and $\beta=20$ provide nearly the same results, suggesting that this local observable is nearly converged in
temperature on the displayed lattice. Its weak size dependence also
supports its local character; it does not establish ground-state
convergence of the longer-range magnetic correlations.

\begin{figure}[t!]
\includegraphics[height=6.0cm,width=7.5cm,angle=0]{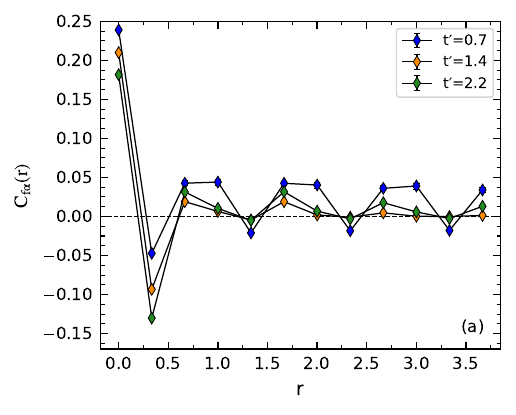}
\includegraphics[height=6.0cm,width=7.5cm,angle=0]{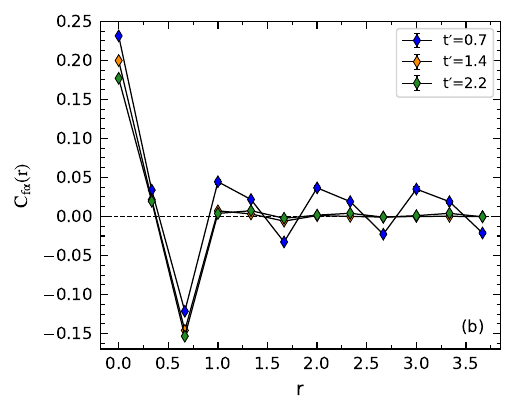} 
\caption{
Real space spin correlations $c_{f\alpha}(r)$ measured from a
decorated $f$ orbital at the origin to the orbitals encountered as one
moves through successive unit cells along the $x$ direction.
Within each unit cell, the horizontal axis steps through the three
inequivalent orbital types in the order $f$, $d$, and $p$, with fractional
offsets used only to separate these orbital positions visually.
Thus, integer values of $r$ correspond to $ff$ correlations in successive
unit cells, while the intermediate points correspond to $fd$ and $fp$
correlations.
Because the $p_x$ and $p_y$ orbitals are symmetry equivalent, and likewise
the $f_x$ and $f_y$ orbitals in $\hat{\cal H}_{{\rm loc},p}$, only 
representative $p$ and $f$ orbitals are shown.
Panel (a) gives the correlations for $\hat{\cal H}_{{\rm loc},d}$, and
panel (b) gives the corresponding results for
$\hat{\cal H}_{{\rm loc},p}$.
The dashed line marks zero correlation.
\label{fig:RealSpaceCorrelation}
}
\end{figure}

The full spatial dependence of the correlation  $c_{f\alpha}(r)$ 
of the spin of an $f$ orbital in a reference unit cell with 
other orbitals $\alpha$ and unit cells $r$,  is shown in 
Fig.~\ref{fig:RealSpaceCorrelation} for both the geometries studied--
$\hat{\cal H}_{{\rm loc},d}$ in panel (a) and
$\hat{\cal H}_{{\rm loc},p}$ in panel (b).
The strong positive value at $r=0$ is the on-site $ff$ correlation, while
the large negative nearest point reflects the anti-correlation between
the decorated $f$ orbital and the backbone orbital to which it is
hybridized (as in Fig.~\ref{fig:singlet}).
The subsequent alternating positive and negative correlations show the
spatial magnetic pattern away from the reference $f$ orbital. 
At $t^{\prime}=0.7$ the values of $c_{f\alpha}(r)$ remain large throughout
the lattice and exhibit the signs expected for ferrimagnetic order.  

At $t'=1.4$, where local antiferromagnetic correlations
between hybridized orbitals are well developed
(Fig.~\ref{fig:singlet}), $c_{f\alpha}(r)$ is small for $r>0$
on the displayed lattice. This suppression is not evidence for an
intermediate singlet ground state: Lieb's theorem requires an
extensive ground-state spin in both geometries.
Instead its origin is 
that the temperature scale for the development
of extended magnetic correlations is particularly small in this
hybridization regime. 

Finally,  at $t^{\prime}=2.2$ there is a suggestion of a revival in 
$c_{f\alpha}(r)$. 
This is especially evident for
$\hat{\cal H}_{{\rm loc},d}$ in panel (a), an effect which will be 
seen more clearly in the structure factors discussed below.

\begin{figure}[t!]
\includegraphics[height=6.0cm,width=7.5cm,angle=0]{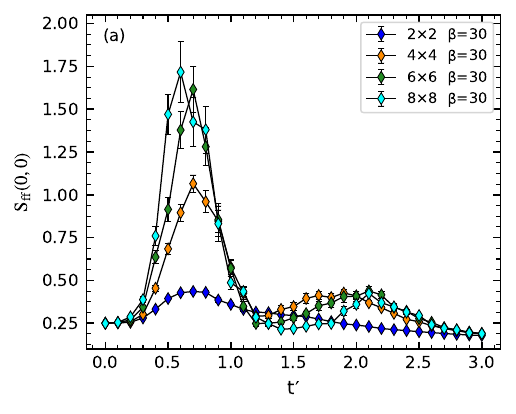}
\includegraphics[height=6.0cm,width=7.5cm,angle=0]{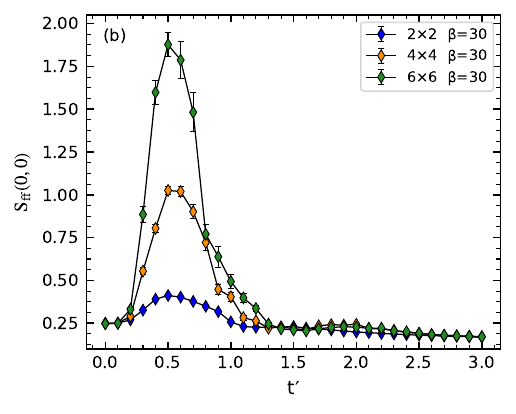} 
\caption{
The ferromagnetic structure factor on the localized $f$ orbitals is shown as a function of $t^{\prime}$
Panel (a) shows $S_{ff}(0,0)$ for the geometry
of $\hat {\cal H}_{{\rm loc},d}$ for four lattice sizes $L= 2, 4, 6, 8$, while panel (b) shows $S_{ff}(0,0)$ for the geometry of $\hat{\cal H}_{{\rm loc},p}$ for three lattice sizes $L= 2, 4, 6$.
The results are shown at inverse temperature $\beta=30$; convergence to the ground state cannot be assumed uniformly across all hybridizations.
\label{fig:SqDD0}
}
\end{figure}

Turning to the structure factor, Fig.~\ref{fig:SqDD0} displays
$S_{ff}(0,0)$ as a function of hybridization $t'$ for
$\hat{\cal H}_{{\rm loc},d}$ (top) and
$\hat{\cal H}_{{\rm loc},p}$ (bottom).
For $t'\sim0.5$ the structure factors are large and increase with
lattice size, consistent with extended magnetic correlations.
Because the $f$ orbitals hybridize locally with the backbone rather
than directly with one another, the conduction orbitals 
mediate effective interactions between them, analogous to RKKY
coupling~\cite{ruderman54,kasuya56,yosida57}.

At small but nonzero hybridization, the magnetic energy scale
may become very small, accounting for the weak correlations at the
finite inverse temperatures shown here; analogous small energy
scales occur in other PAM geometries~\cite{vekic95}. At intermediate
hybridization the $f$-orbital structure factors are strongly
suppressed. This cannot be identified with a spin-singlet ground
state, because the sublattice imbalance fixes a nonzero
$S_{\rm GS}$ for all $t'>0$. Instead, the data suggest a marked
reduction in extended magnetic correlations at the temperatures
studied, possibly associated with a minimum in their characteristic
energy scale. A secondary maximum around $t'\sim2$ is especially
visible in the four-orbital geometry. We next compare it with the
corresponding response on the backbone orbitals.

\begin{figure}[t!]
\includegraphics[height=6.0cm,width=7.5cm,angle=0]{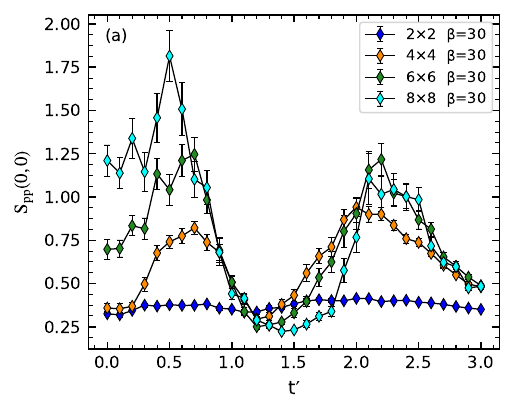}
\includegraphics[height=6.0cm,width=7.5cm,angle=0]{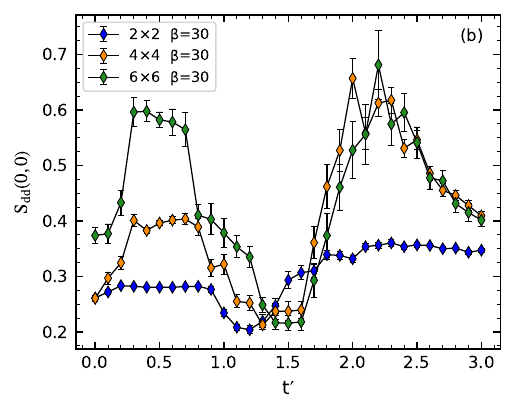} 
\caption{
The ferromagnetic structure factor on the undecorated Lieb-backbone orbitals.
Panel (a) shows $S_{pp}(0,0)$ for $\hat{\cal H}_{{\rm loc},d}$, while panel (b) shows $S_{dd}(0,0)$ for $\hat{\cal H}_{{\rm loc},p}$.
The inverse temperature is $\beta=30$.
\label{fig:SqUndecorated}
}
\end{figure}

\begin{figure}[t!]
\includegraphics[height=6.0cm,width=7.5cm,angle=0]{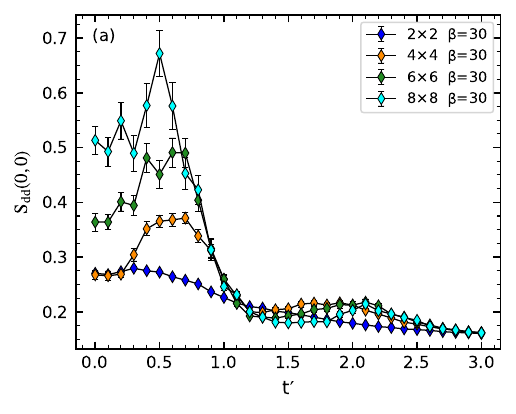}
\includegraphics[height=6.0cm,width=7.5cm,angle=0]{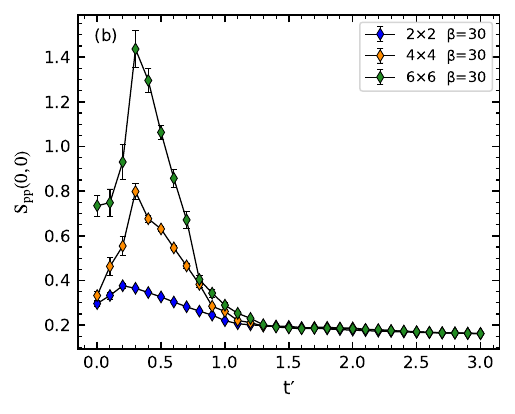} 
\caption{
The ferromagnetic structure factor on the decorated Lieb-backbone orbitals.
Panel (a) shows $S_{dd}(0,0)$ for
$\hat{\cal H}_{{\rm loc},d}$, while panel (b) shows
$S_{pp}(0,0)$ for
$\hat{\cal H}_{{\rm loc},p}$.
The inverse temperature is $\beta=30$.
\label{fig:SqDecorated}
}
\end{figure}

Figures \ref{fig:SqUndecorated} and \ref{fig:SqDecorated}
compare the magnetic correlations of the undecorated and decorated
Lieb-backbone orbitals respectively. 
Notice that in Fig~\ref{fig:SqUndecorated}(a) we show $S_{pp}$ for
$\hat{\cal H}_{{\rm loc},d}$ because there the $p$ orbitals have no localized $f$ orbital partners.
Meanwhile in Fig~\ref{fig:SqUndecorated}(b) we show $S_{dd}$ for
$\hat{\cal H}_{{\rm loc},p}$ because there the $d$ orbitals have no localized $f$ orbital partners.
Similarly, for Fig.~\ref{fig:SqDecorated} the structure factors shown are for orbitals with
localized $f$ partners.
The strongest non-monotonic response is
seen on the undecorated orbitals in Fig.~\ref{fig:SqUndecorated}:
large structure factors at $t'\sim0.5$ are followed by a deep
minimum around $t'\sim1.2$--$1.5$ and a pronounced recovery near
$t'\sim2.2$. 

The recovery can be understood qualitatively in terms of
selective hybridization. At large $t'$, the decorated backbone
orbitals develop strong local antiferromagnetic correlations with
the attached $f$ orbitals, while the undecorated backbone sites
retain low-energy magnetic degrees of freedom. Effective couplings
between these residual degrees of freedom may sustain substantial
spatial correlations even as the local $d$--$f$ or $p$--$f$
anticorrelations strengthen. The microscopic dependence of these
effective couplings on $t'$ requires further analysis.

A related persistence of magnetic correlations at large hybridization
was observed for regular depletion of localized orbitals on a square
lattice~\cite{Costa2018}. The distinguishing observation here
is the pronounced intervening minimum in orbital-resolved structure
factors, which is not comparably prominent in Ref.~\cite{Costa2018}.
Moreover, the two Lieb-lattice decorations give different flat-band
spectra and different sublattice imbalances, allowing the magnetic
response of inequivalent backbone orbitals to be compared directly.

\section{IV.~Conclusions}

We have studied magnetic correlations in two interacting
Lieb-lattice Anderson models distinguished by whether localized
orbitals attach to the $d$ or $p$ sites of the conduction backbone.
Although the local spin anticorrelations on hybridized pairs
increase relatively smoothly with $t'$, the extended magnetic
correlations exhibit a strikingly nonmonotonic dependence on
hybridization. In particular, the undecorated backbone orbitals
show a deep minimum at intermediate $t'$ and substantially
stronger correlations again at larger $t'$.

The recovery of magnetism at large hybridization has a
precedent in depleted periodic Anderson models~\cite{Costa2018};
our main distinction is the intermediate suppression, together with
the orbital-resolved comparison of two different flat-band
geometries. The minimum does not establish an intervening spin-liquid
or singlet ground-state phase. Indeed, Lieb's theorem fixes an
extensive, nonzero total ground-state spin at all $t'>0$ for the
half-filled interacting bipartite Hamiltonians considered here.
Our results instead suggest that the characteristic energy scale for
building extended correlations may be especially small at
intermediate hybridization. 

Although structure factors grow with system size in parts of
the displayed parameter range, finite-size extrapolations
of the orbital-resolved order parameters are small or compatible with
zero. 
Nevertheless, our calculations establish there is a large and highly orbital-dependent
nonmonotonic magnetic response.

It would be interesting to investigate whether the nonmonotonic behavior observed here could also be seen in experiments of Lieb-lattice materials. Specifically, decorating the $d$ and $p$ sites with additional orbitals would allow the hybridization $t'$ to be varied independently of the backbone hopping. Experiments could then test whether increasing $t'$ weakens and subsequently strengthens magnetic correlations on undecorated sites while the spins on hybridized pairs become increasingly anti-aligned. Ultracold atom experiments have already demonstrated ferrimagnetic correlations on the Lieb lattice~\cite{Lebrat2026Ferrimagnetism}-- studying additional orbitals requires further experimental precision and control.

Concerning future directions,
the effect of depletion of localized $f$ orbitals 
on the NMR coherence temperature $T_{*}$
has been an area of interest~\cite{mendes2017impurities,costa2019coherence}.
Such work was done for `conventional' backbone geometries with dispersive bands
such as the square lattice.
The flat band present in the
models here might affect that phenomenology, including the linear suppression
of $T_{*}$ with impurity concentration.  However, this requires measurement of the
imaginary-time displaced magnetic structure factor which is more computationally
intense than the equal time observables obtained in the present work.

\vskip0.10in
\noindent
\underbar{Acknowledgements:}
The work of AA and RTS was supported by 
grant DOE DE-SC0014671 funded by
the U.S. Department of Energy, Office of Science. 
The work of RH was supported by The Glynn Family Honors Program at the University of Notre Dame.
Parts of the computations of this work were performed at the Notre Dame Center for Research Computing.


\bibliography{decoratedlieb}

\end{document}